\documentstyle[12pt,amssymbols]{article}
\def\a{\alpha}

\def\d{\delta}
\def\l{\lambda}
\def\m{\mu}

\def\rmat{R- matrix}
\def\L{L_{n,a}(\lambda)}
\def\n{\nu_a}
\def\h{h_n}

\def\rep{representation }
\def\reps{representations }

\def\l{\lambda}
\def\bp{U_q(b_+)}
\def\bm{U_q(b_-)}
\def\aux{ auxiliary }

\def\RMAT{R=\sum_i A_i\otimes B_i \in U_q(b_+)\otimes U_q(b_-) }

\def\half{{1\over2}}
\def\sl{U_q(\widehat{sl}_2)}
\def\s{U_q(sl_2)}
\def\<{\langle}
\def\>{\rangle}

\newcommand{\bn}{\begin{equation}}
\newcommand{\ed}{\end{equation}}
\newcommand{\fmt}{ fundamental $R$--matrix }
\newcommand{\A}{{\cal A }}
\newcommand{\Rmt}{ universal $R$--matrix }
\newcommand{\hsp}{\mbox{$\hspace{.5in}$}}

\begin{document}
\rightline{LPTHE-96-28}
\vspace{1.7in}
\centerline{\Large {\bf Universal R-matrix and Quantum Volterra Model }}
\vskip 1.4cm
\centerline{\large Alexander Antonov
\footnote[1]{e-mail address: antonov@lpthe.jussieu.fr and 
antonov@landau.ac.ru}}
\vskip1.2cm
{\it\centerline{  Laboratoire de Physique Th\'eorique et Hautes
Energies}
\centerline{ Universit\'e Pierre et Marie Curie, Tour 16 1$^{er}$
		\'etage, 4 place Jussieu}
\centerline{75252 Paris cedex 05-France}
\centerline{and}
\centerline{ Landau Institute for Theoretical Physics }
\centerline{Kosygina 2, GSP-1, 117940 Moscow V-334, Russia }}
\vskip1cm
\centerline{\bf June 1996}
\vskip4cm

\centerline{\bf Abstract}
\noindent
In this paper
we explicitly prove that Integrable System solved by
Quantum Inverse Scattering Method can be described with the 
pure algebraic object (Universal R-matrix) and proper
algebraic \reps. Namely, on the example of the Quantum Volterra model
we construct
L-operator and \fmt from \Rmt for Quantum Affine $\sl$ Algebra
and q-oscillator \rep for it. In this way there exists an
equivalence between the Integrable System with symmetry
algebra \A and the \rep of this algebra.

\vspace{2in}
\setcounter{section}{-1}

\section{Introduction}
The problem of diagonalization of an infinite set of mutually commuting 
Integrals of Motion (IM) for some Integrable Quantum Theory is solved
by the Quantum Inverse Scattering Method (QISM) \cite{FST}.
The generating function for IM is trace of monodromy matrix $T(\l)$
depending on some spectral parameter.

Another approach for the objects of QISM is based on so called \Rmt. 
An explicit formula 
for \Rmt for quantum affine 
algebras was found in \cite{hor}.

For some algebra $\A$ the \Rmt lies in the square 
$R\in \A\otimes \A$ and satisfies the Yang-Baxter (YB) eq.
$$
R_{12} R_{13} R_{23} = R_{23} R_{13} R_{12}, 
$$
in the $\A\otimes \A \otimes \A$.
We write $R=\sum_i A_i \otimes B_i \in \A\otimes  \A$ and let  
$$
R_{12}=\sum_i
A_i \otimes B_i \otimes 1 ,\,\,
R_{13}=\sum_i A_i  \otimes 1 \otimes B_i ,\,\,
R_{23}=\sum_i  1 \otimes A_i \otimes B_i . 
$$

Now we formulate the
{\bf Main Proposition}.
 
{\it The Integrable System with symmetry algebra $\A$ solved by QISM can be described using the \Rmt $R\in \A\otimes \A$ 
for algebra $\A$ and its proper \reps.
In fact, the \rep of the first algebra $\A$ (quantum \rep) is connected 
with the  Integrable System
in consideration and the \rep of the second one (\aux \rep) 
establishes the sort 
of the QISM object (L-operator, \fmt, Baxter Q-operator etc.)
}

First, it was formulated in \cite{SB}.

Later we will consider only the case of $\sl$ algebra.
Monodromy matrix can be got from the \Rmt
$R\in \sl\otimes \sl$ using 
some quantum representation
(depending on a integrable model in consideration)
for first $\sl$ algebra and \aux
matrix $n\times n$ representation 
with a spectral parameter
for the second $\sl$ algebra in the square  $\sl\otimes\sl$.

Another example deals with the Baxter  operator 
$Q(\l)$ satisfying Baxter eq. \cite{Bax}
\bn
T(\l) \, Q(\l)= Q(q^2 \l)+ Q(q^{-2} \l)
\label{TBeq}
\ed
We have Baxter operator from the \Rmt
$R\in \sl\otimes \sl$ using 
some quantum representation
for the first $\sl$ algebra and \aux
infinite dimensional q-oscillator representation 
\cite{BLZ1} 
for the second $\sl$ one in the square  $\sl\otimes\sl$.

According to the methods of \Rmt for monodromy matrix or Q- operators
we specify representation only of the {\it second} $\sl$ algebra
for  $R\in \sl \otimes  \sl$, and representation of the first one
(depending on a concrete model in hand) is not fixed.
So application of \Rmt gives us {\it universal} technique for deriving
algebraic relations between different components of the QISM 
\cite{AF}.

{\it In this paper we consider the infinite dimensional q-oscillator
\rep as a quantum one for \Rmt. It occurs that an integrable model described
by this \rep is Quantum Volterra Model} \cite{V}, \cite{B}. 
We derive all the Volterra QISM objects, such as L-operator and
\fmt \cite{V} from \Rmt using q-oscillators as quantum \rep.

The text is organized as follows.
In {\bf Section 1} we remind some key moments of QISM and interpret
monodromy matrix via \Rmt representations. 
We give the explicit construction of the Volterra L-operator
in terms of \Rmt in
{\bf Section 2}.
In {\bf Section 3} we derive Volterra \fmt from \Rmt and explain its position
in YB eq. The discussion and conclusion
are presented in {\bf Section 4}.

\vspace{0.2in}
{\bf Acknowledgments.}

\noindent
The author is thankful to L.Baulieu and
Laboratoire de Physique Th\'eorique et Hautes \'Energies,
Paris, France for the hospitality.
The author would like to thank A.Belavin, B.Enriquez, L.Faddeev,
B.Feigin, 
S.Khoroshkin, F.Smirnov,
A.Volkov
for illuminating discussions.


\section{ Origin of QISM Ingredients from Universal R-matrix }
In this section we use the notations from \cite{SB}.
The quantum L-operator is the main object of the QISM.
$\L$ is a matrix in an auxiliary space $\n$, the matrix 
elements of $\L$
are operators in a Hilbert space $\h$, 
$n=1, \cdots , N $
associated with the site of the lattice, depending on the 
spectral parameter
$\l$.
The operators in different sites commute.
In this way, the operator $\L$ is the operator in the square 
 $\h\otimes\n$.

The fundamental commutation relations for the matrix 
elements of $\L$
is the YB eq.:
\begin{equation}
L_{n,a_1}(\mu) L_{n,a_2}(\lambda)R_{a_1,a_2}(\lambda / \mu)= 
R_{a_1,a_2}(\lambda / \mu) L_{n,a_2}(\lambda) L_{n,a_1}(\mu) 
\label{fcr}
\end{equation}
This is an equation in $ h_n\otimes\nu_{a_1} \otimes \nu_{a_2}$. 
The indices $a_1$ and
$a_2$ and the variables $\lambda$ and $\mu$ are associated with
the auxiliary spaces $\nu_1$ and $\nu_2$, respectively.
The matrix $R_{a_1,a_2}$ is one in the space 
$\nu_{a_1} \otimes \nu_{a_2} $.

Further, we consider only the special form of $R_{a_1,a_2}$- matrix, 
so called
trigonometric R-matrix, $4 \times 4$. A lot of interesting 
models on a lattice
(XXZ spin model, the lattice Sine- Gordon system, 
Volterra system etc.)
and in continuum  
are described by the 
trigonometric R-matrix.   

Another important object of the QISM is the monodromy matrix
\begin{equation}
M_a(\lambda) = L_{N,a}(\lambda) L_{N-1,a}(\lambda) \cdots  L_{1,a}(\lambda).
\label{tt}
\end{equation}
The monodromy matrix satisfies commutation relations identical 
with the  YB ones
for L- operators (\ref{fcr}). It follows that the traces 
$T_a(\l)=\hbox {tr}_{\n} M_a (\l)$ over the auxiliary spaces
of the monodromy 
matrix with the different parameters commute 
$[T_{a_1}(\l), T_{a_2}(\mu)]=0$.
That is why the trace of the monodromy matrix is the 
generating function for the
integrals of motion.

It is useful to consider so called fundamental 
R-matrix \cite{FTT},
$L_{n_1,n_2}(\l)$, i.e. the operator in $h_{n_1}\otimes h_{n_2}$. 
In other words the auxiliary space coincides
with the quantum one.
The trace of the monodromy matrix for the fundamental R-matrix 
gives the set of {\it local}
integrals of motion.

It is possible to describe all the objects of QISM 
(trigonometric \rmat,
L- operator and monodromy matrix as well as 
fundamental R-matrix)
with the only algebraic structure--- \Rmt.

We will briefly remind some facts about it \cite{hor}. 
Below, we consider
the simple case of the affine quantum $\sl$ algebra 
with Cartan matrix 
 $A=(a_{ij})$, $i,j=0,1$,
$
A = \left( \begin{array}{cc}
2 & -2 \\
-2 &  2
\end{array}
\right) $.
This is an associative
algebra with generators 
$e_{\pm \alpha_{i}},\; k_{\alpha_{i}}^{\pm 1}(=q^{\pm h_{\a_i}}),\;
 (i=0,1)$, and the defining relations
$$
[k_{\alpha_{i}}^{\pm 1}, k_{\alpha_{j}}^{\pm 1}]=0,
\hspace{.5in}
k_{\alpha_{i}}e_{\pm \alpha_{j}}= q^{\pm (\alpha_{i},\alpha_{j})}
e_{\pm\alpha_{j}} k_{\alpha_{i}},
$$
$$
[e_{\alpha_{i}}, e_{-\alpha_{j}}] = \delta_{ij}\frac{k_{\alpha_{i}} -
k_{\alpha_{i}}^{-1}}{q-q^{-1}},
$$
$$
(ad_{q'}e_{\pm\alpha_{i}})^{1-a_{ij}} e_{\pm\alpha_{j}}=0 
\hspace{.5in}
\hbox {for}\;\,\,\, i\neq j, \:q'=q,q^{-1},
$$
where $(ad_{q}e_{\alpha})e_{\beta}$ is a q-commutator:
$$
(ad_{q}e_{\alpha})e_{\beta}\equiv
{[e_{\alpha},e_{\beta}]}_{q}=e_{\alpha}e_{\beta}-q^{(\alpha,\beta )}
e_{\beta}e_{\alpha}
$$
and  $(\alpha ,\beta )$ is a scalar  product of the roots $\alpha$
and $\beta$: $(\alpha_{i},\alpha_{j})=a_{ij}$.
The condition $ 1=k_{\a_0} k_{\a_1}$ is also imposed.

We define a comultiplication in $\sl$ by the formulas
$$
\Delta (k_{\alpha_{i}})=k_{\alpha_{i}}\otimes k_{\alpha_{i}},
$$
$$
\Delta (e_{\alpha_{i}}) = e_{\alpha_{i}}\otimes
k_{\alpha_{i}} + 1 \otimes e_{\alpha_{i}}
\,\,\,\,
\Delta (e_{-\alpha_{i}}) = e_{-\alpha_{i}}\otimes 1+
k^{-1}_{\alpha_{i}}\otimes e_{-\alpha_{i}},
$$
and the twisted comultiplication
$$
\Delta '(k_{\alpha_{i}})=k_{\alpha_{i}}\otimes k_{\alpha_{i}},\;
$$
$$
\Delta' (e_{\alpha_{i}}) =k_{\alpha_{i}}\otimes e_{\alpha_{i}} + 
e_{\alpha_{i}}\otimes 1,
\,\,\,\,
\Delta ' (e_{-\alpha_{i}}) = 1\otimes e_{-\alpha_{i}}+
 e_{-\alpha_{i}}\otimes  k^{-1}_{\alpha_{i}},
$$

By the definition, the \Rmt is an object $R$ in $\sl\otimes\sl$ such that
$\Delta '(g) R = R \Delta (g)$, for any $g \in \sl$ and 
\begin{equation}
R_{12} R_{13} R_{23} = R_{23} R_{13} R_{12}, 
\label{uyb}
\end{equation}
which is the universal form of the YB eq. in the $\sl\otimes\sl\otimes\sl$.
We write $R=\sum_i A_i \otimes B_i \in \sl\otimes \sl$ and let  
$$
R_{12}=\sum_i
A_i \otimes B_i \otimes 1 ,\,\,
R_{13}=\sum_i A_i  \otimes 1 \otimes B_i ,\,\,
R_{23}=\sum_i  1 \otimes A_i \otimes B_i . 
$$

Along with the commutation relations for $\sl$ we need 
an antiinvolution  $(^{*})$ ,
defined as
$(k_{\alpha_{i}})^{*}=k_{\alpha_{i}}^{-1},\;$
$(e_{\pm\alpha_{i}})^{*}=e_{\mp\alpha_{i}},\;$
$(q)^{*}=q^{-1}$.

We use also the following standard notations for q-exponent
and q-numbers
$$
\hbox{exp}_{q}(x) := 1 + x + \frac{x^{2}}{(2)_{q}!} + \ldots +
\frac{x^{n}}{(n)_{q}!} + \ldots = \sum_{n\geq 0} \frac{x^{n}}{(n)_{q}! },$$

$$(a)_{q}:=\frac{q^{a}-1}{q-1},\,\,\,\,\,\,\,\,\,\,\,\,
[a]_{q}:=\frac{q^{a}-q^{-a}}{q-q^{-1}}$$

Now we define the Cartan-Weyl generators of the $\sl$.
Let $\alpha_1$ and $ \alpha_0=\delta -\alpha_1$ 
are simple roots for the affine
algebra $\widehat{sl_{2}}$ then $\delta=\alpha_1+\beta$ is a minimal
imaginary root.  We fix the following normal ordering in the system of the
positive roots: \
$$
\alpha_1,\;\alpha_1+\delta, \alpha_1+2\delta\ldots ,
\delta,\; 2\delta, \ldots , \ldots ,
\a_0 +2\d ,\a_0 +\d ,\a_0 \ .
$$
We put
$$
e'_{\delta}=e_{\delta}=
[e_{\alpha_1},e_{\a_0}]_{q} \ ,
$$
$$
e_{\alpha_1+l\delta}=(-1)^{l}
\left( [2]_{q}\right)^{-l}
(ad\ {e'}_{\delta})^{l}e_{\alpha_1} \ ,
$$
\bn
e_{\a_0+l\delta}=
\left( [2]_{q}\right)^{-l}
(ad\ {e'}_{\delta})^{l}e_{\a_0} \ ,
\label{gener}
\ed

$$
{e'}_{l\delta}=
[e_{\alpha_1+(l-1)\delta},e_{\a_0}]_{q}.
$$

Let $E(z)$ and ${E'}(z)$ be the generating functions for
$e_{n\d}$ and for ${e'}_{n\d}$:
\bn
E(z)=\sum_{n \geq 1}e_{n\d}z^{-n},
\hsp
{E'}(z)=\sum_{n \geq 1}{e'}_{n\d}z^{-n},
\label{transf}
\ed
connected via
$$
(q-q^{-1})E(z)= \log \left( 1+(q-q^{-1}){E'}(z)\right)
$$
The negative root vectors are given by the rule
$e_{-\gamma}$ = $e_{\gamma}^{*}$.

The universal $R$-matrix for $U_q(\widehat{sl}_2)$ has the following form
 \cite{hor}:
\[
{\cal R}=\left(\prod_{n\geq 0}^{\rightarrow} \exp_{q^{-2}}
\left( (q-q^{-1})e_{\alpha_1+n\delta} \otimes e_{-\alpha_1-n\delta}
\right)\right)\cdot
\]
\[
\exp\left( \sum_{n>0}(q-q^{-1})\frac{n(e_{n\delta}
\otimes e_{-n\delta})}{[2n]_q}\right)\cdot
\]
\bn
\left(\prod_{n\geq 0}^{\leftarrow}\exp_{q^{-2}}\left( (q-q^{-1})
e_{\a_0+n\delta}\otimes e_{-\a_0-n\delta}\right)\right)\cdot {\cal K}
\label{urm}
\ed
where the order of $n$ is direct in the first product and it is inverse
in the second one. Factor ${\cal K}$ is defined by the  formula:
$$
{\cal K}=q^{{\frac{h_\a\otimes h_\a}{2}}}
$$

We know (\ref{urm}) that \Rmt belongs to the square of $\bp\otimes\bm$,
where $b_\pm$ are positive (negative) borel subalgebras of $\sl$,
generated by 
$e_{ \alpha_{i}},\; k_{\alpha_{i}}^{\pm 1}$
and
$e_{ -\alpha_{i}},\; k_{\alpha_{i}}^{\pm 1}
,\; (i=0,1)$, respectively.

One can present QISM L-operators from the \Rmt using the certain
representation of $\sl$. For example, the trigonometric $4\times 4$ \rmat
is got from the \Rmt representing $\sl$ in terms of matrix $2\times 2$
with the spectral parameter.

It is possible to get the $\L$ operator from the \Rmt 
$R\in \bp\otimes\bm$ representing 
the first algebra $\bp$
in quantum (infinite dimensional) space and
the second one $\bm$ in matrixes with the spectral parameter (\aux 
representation).
Substituting the comultiplied $\Delta^{(N-1)}\sl$ quantum space 
generators in the
$\bp$ part of the \Rmt we get the monodromy matrix of N-sites (\ref{tt}).

Now we illustrate the YB eq. (\ref{fcr})
from the point of view of the universal one (\ref{uyb}). Eq. (\ref{uyb})
is one for the algebraic elements in $\sl\otimes\sl\otimes\sl$.
To get the eq.  (\ref{fcr}) we represent 
the first algebra in quantum space $\h$, 
the second and the third one ---
in finite dimensional (with respect to $\s$) representations, say 
matrix with the spectral parameter.


\section{Explicit Construction of Volterra L-operator
from Universal R-matrix }

Quantum Volterra L-operator
\bn
L(\l)=
 \left( \begin{array}{cc}
u & -\l v^{-1}\\
\l v & u^{-1}
\end{array}
\right) 
\label{vlop}
\ed
was introduced in \cite{V}
using the Weyl pair $u$ and $v$
\bn
uv=q vu
\label{wp}
\ed
Volterra L-operator satisfies YB eq. (\ref{fcr}) for the trigonometric
R-matrix. To derive L-operator (\ref{vlop}) from the \Rmt
we will use the q-oscillator representation.
The importance of this \rep was pointed out
in \cite{BLZ1}. 

It was shown that  one can get Baxter
$Q(\l)$ operator
from  $\RMAT$ by the following \aux q-oscillator \rep $V_\pm$ 
of the second algebra $\bm$ 
$$
e_{\a_0} e_{\a_1}- q^2 e_{\a_1} e_{\a_0}= \frac {\l^2}{q^{-2}-1}
\hspace{.5in}
(\hbox{the representation } V_+(\l))
$$

and
 
$$
e_{\a_1} e_{\a_0}- q^2 e_{\a_0} e_{\a_1}= \frac {\l^2}{q^{-2}-1},
\hspace{.5in}
(\hbox{the  representation } V_-(\l)),
$$

we have
$$
Q_+(\l)=\hbox{tr}_{ V_+(\l) } R
\hspace{.5in}
Q_-(\l)=\hbox{tr}_{ V_-(\l) } R
$$
for {\it any} quantum representation  of $\bp$ (for {\it any} integrable model)
\footnotemark
\footnotetext[1]
{We choose the infinite dimensional \rep space $V_\pm (\l)$ such that the trace
$Q_\pm(\l)=\hbox{tr}_{ V_\pm(\l) } R$  exists.}.
Such Q- operators satisfies Baxter eq. (\ref{TBeq}).

Here, q-oscillator \rep (as \aux one) describes {\it the sort} of the QISM
object: it gives Baxter Q-operator for any quantum \rep of $\bp$.

Now natural question arises:

{\it Which Integrable Model corresponds to q-oscillator \rep
as a quantum one, i.e. the \rep of $\bp$ algebra for $\RMAT$?}

We will prove that this system is the Quantum Volterra Model.   
For constructing Volterra L-operator we use the special case
 of  $V_\pm (\l)$ \reps. Namely, define  $\bp$ algebra \rep $W(\l)$
as follows
\bn
e_{\a_1}=\l e (q-q^{-1})^{-1},
\hsp 
e_{\a_0}=\l e^{-1} (q-q^{-1})^{-1},
\hsp
k_{\a_1}=k(=q^h)
\label{repw}
\ed
where $ke=q^2 ek$. (Representation
 $V_\pm(\l)$ degenerates in commuting 
$e_{\a_1}$ and $e_{\a_0}$ generators).

According to the definition of algebra generators (\ref{gener})
we have
$$
e'_\d=\frac{\l^2}{q^2-1}
$$
Using the transformation (\ref{transf})
one obtains
\bn
e_{n\d}=(-)^{n+1}\frac{q^{-n}}{q-q^{-1}}\frac{\l^{2n}}{n}
\label{edelta}
\ed
with the other $\bp$ generators 
\bn
e_{\alpha_1+l\delta}=
e_{\a_0+l\delta}=0\,\,\,
\hbox{for}\,\,\, l=1,2,...
\label{degen}
\ed

For constructing L-operator for Volterra model
consider the evaluation representation for the $\sl$ in terms of 
$\s\otimes \Bbb {C}[\l,\l^{-1}]$
\begin{equation}
e_{\pm \alpha_{1}}=  \l^{\pm 1} e_{\pm \alpha},
\hspace{.5in}
e_{\pm \alpha_{0}}= -\l^{\pm 1} e_{\mp \alpha},
\hspace{.5in}
k_{\alpha_{1}}= k_\a,
\label{ev}
\end{equation}
where $\s$ algebra has generators $ e_{\pm \alpha} $ and $k_\a^{\pm 1}$
with the ordinary commutation relations
$$
k_\a e_{\pm \alpha}= q^{\pm 2}
e_{\pm\alpha} k_\a,
\hspace{.5in}
[e_{\alpha}, e_{-\alpha}] = \frac{k_\a - k_\a^{-1}}{q-q^{-1}},
$$

Introduce the irreducible $2$ -dimensional matrix representation of $\s$.
\bn
e_\a = \left( \begin{array}{cc}
0 & 1 \\
0 & 0
\end{array}
\right) 
,\,\,\,
e_{-\a}= \left( \begin{array}{cc}
0 & 0 \\
1 & 0
\end{array}
\right) 
,\,\,\,
k_\a= \left( \begin{array}{cc}
q & 0 \\
0 & q^{-1}
\end{array}
\right)
\label{repmat}
\ed
One can define the evaluation representation of the affine $\sl$ algebra 
from (\ref{ev})
and  the \rep (\ref{repmat}) of the simple $\s$ algebra. 
For the generators we have
\bn
e_{\a_1 +m\d}=\l^{2m+1}k_\a^{-m}e_\a,
\hsp
e_{\a_0 +m\d}=-\l^{2m+1} e_{-\a} k_\a^{-m},
\label{polger}
\ed

and also
\bn
e_{-(\a_1 +m\d)}= \l^{-2m-1}e_{-\a}k_\a^m,
\hsp
e_{-(\a_0 +m\d)}=- \l^{-2m-1} k_\a^me_\a
\label{otger}
\ed
and for imaginary root vectors
\bn
e_{m\d}= - \frac{q^{ m}+q^{-3m}-k_\a^{-m}-q^{-2m}k_\a^{-m}}
{m(q-q^{-1})}\l^{2m}
\label{polmnim}
\ed

\bn
e_{-m\d}= \frac{q^{- m}+q^{3m}-k_\a^{m}-q^{2m}k_\a^{m}}
{m(q-q^{-1})}\l^{-2m}
\label{otmnim}
\ed

Now we are in position to derive Volterra L-operator from \Rmt.
Let us denote by $L_\half(\l)$ some L-operator got
from \Rmt $\RMAT$ by following representation
\begin{itemize} 
\item the first algebra $\bp$ of the \Rmt is represented in
quantum space $W(\l)$;
\item for the second algebra $\bm$ we take the \aux evaluation representation 
of $\s$ spin $ \half $. 
\end{itemize}

The first and the third factors in formula (\ref{urm})
can be evaluated easily
$$
\left(\prod_{n\geq 0}^{\rightarrow} \exp_{q^{-2}}
\left( (q-q^{-1})e_{\alpha_1+n\delta} \otimes e_{-\alpha_1-n\delta}
\right)\right)=
1+ (q-q^{-1})e_{\alpha_1} \otimes e_{-\alpha_1}=
 \left( \begin{array}{cc}
1 & 0\\
\l e & 1
\end{array}
\right) 
$$

and

$$
\left(\prod_{n\geq 0}^{\leftarrow}\exp_{q^{-2}}\left( (q-q^{-1})
e_{\a_0+n\delta}\otimes e_{-\a_0-n\delta}\right)\right)=
1+ (q-q^{-1})e_{\alpha_0} \otimes e_{-\alpha_0}=
 \left( \begin{array}{cc}
1 & -\l e^{-1}\\
0 & 1
\end{array}
\right)
$$
because of the property (\ref{degen}).

Using the formulae (\ref{edelta}) and  (\ref{otmnim}) we have the second
factor (up to a constant one)
$$
\exp\left(\sum_{n>0}(q-q^{-1})\frac{n(e_{n\delta}
\otimes e_{-n\delta})}{[2n]_q}\right)
=\exp\left((q-q^{-1}) \sum_{n>0} (-)^n \frac{\l^{2n}}{n}
\right)
\left( \begin{array}{cc}
1 & 0\\
0 & 0
\end{array}
\right)=
\left( \begin{array}{cc}
\frac{1}{1+\l^2} & 0 \\
0 & 1
\end{array}
\right)
$$
The factor ${\cal K}$ reduces to
$$
{\cal K}=q^{{\frac{h_\a\otimes h_\a}{2}}}=
\left( \begin{array}{cc}
k^\half & 0\\
0 & k^{-\half}
\end{array}
\right)
$$
Finally, one gets the expression for $L_\half(\l)$ operator
\bn
L_\half(\l)=
\left( \begin{array}{cc}
k^\half & -\l e^{-1} k^{-\half} \\
\l e k^\half & k^{-\half}
\end{array}
\right)
\label{lop}
\ed

Choosing 
$$
e=v u^{-1}
\,\,\,
\hbox{and}
\,\,\,
k=u^2
$$
we see that $L_\half(\l)$ coincides with Volterra L-operator (\ref{vlop}).

\section{Fundamental R-matrix for Volterra Model}
In this chapter we will construct Volterra model \fmt  $R_f$,
derived in \cite{FZ} and \cite{V}. 
Introduce $\bm$ algebra \rep $\tilde W(\l)$
as follows
\bn
e_{-\a_1}=\l^{-1} \tilde e^{-1} (q-q^{-1})^{-1},
\hsp 
e_{-\a_0}=\l^{-1} \tilde e (q-q^{-1})^{-1},
\hsp
k_{\a_1}=\tilde k(=q^{\tilde h})
\label{repwm}
\ed
(where $\tilde k \tilde e=q^2 \tilde e \tilde k$),
with the other $\bm$ generators 
\bn
e_{-\alpha_1+l\delta}=
e_{-\a_0+l\delta}=0,\,\,\, e_{-(l+1)\delta}=const\,\,\,
\hbox{for}\,\,\, l=1,2,...
\label{degenm}
\ed

To get  $R_f$ from \Rmt $\RMAT$ one needs to represent
$\bp$ algebra in quantum space $W(\l)$
(\ref{repw})-(\ref{degen})
and $\bm$ algebra in space $\tilde W(\l)$
(\ref{repwm})-(\ref{degenm}). According to the definition of
the \fmt the quantum space coincides with the \aux one.

In this way the first and the third factors in formula (\ref{urm})
can be evaluated easily
$$
\left(\prod_{n\geq 0}^{\rightarrow} \exp_{q^{-2}}
\left( (q-q^{-1})e_{\alpha_1+n\delta} \otimes e_{-\alpha_1-n\delta}
\right)\right)=\exp_{q^{-2}}
\left( (q-q^{-1})^{-1}e \otimes  \tilde e^{-1} \l
\right)
$$

and

$$
\left(\prod_{n\geq 0}^{\leftarrow}\exp_{q^{-2}}\left( (q-q^{-1})
e_{\a_0+n\delta}\otimes e_{-\a_0-n\delta}\right)\right)=
\exp_{q^{-2}}\left(  (q-q^{-1})^{-1} 
e^{-1}\otimes \tilde e \l \right)
$$
because of the statements (\ref{degen}) and  (\ref{degenm}).
We see that imaginary roots for both $\bp$ and $\bm$ \reps
are some constants, so calculating \fmt up to a constant factor
one can omit the exponent containing imaginary roots.

Finally, we have
\bn
R_f(\l)=\exp_{q^{-2}}
\left( (q-q^{-1})^{-1}e \otimes \tilde e^{-1} \l
\right)\cdot
\exp_{q^{-2}}\left( (q-q^{-1})^{-1}
e^{-1}\otimes \tilde e \l \right)\cdot
q^{{\frac{h\otimes \tilde h}{2}}}
\label{frm}
\ed

Now we show the place of the \fmt in YB eq.
The universal form of YB eq. reads
\footnotemark
\footnotetext[2]
{Note that it is impossible to enlarge the \rep  $W(\l)$ from
$\bp$ to the whole algebra $\sl$.}
$$
R_{12} R_{13} R_{23} = R_{23} R_{13} R_{12}\in
\bp\otimes\sl\otimes\bm
$$
Here for the first $\bp$ algebra we choose the \rep $W(\l)\equiv n_1$,
for the second one $\sl$ --- the evaluation \rep $\equiv a$
and for the third one --- the \rep $\tilde W(\l)\equiv n_2$.

In this way one has
\begin{equation}
L_{n_1,a}(\l) R_{n_1,n_2}(\mu\l) L_{a,n_2}(\mu)=
L_{a,n_2}(\mu) R_{n_1,n_2}(\mu\l) L_{n_1,a}(\l), 
\label{fyb}
\end{equation}
where $L_{n_1,a}(\l)$ is the Volterra L-operator and
$ R_{n_1,n_2}(\l)\equiv R_f (\l)$ is the Volterra \fmt.

Let us denote the trace of monodromy matrix for \fmt as follows
$$
I(\lambda) = \hbox{tr}_h R_{N,h}(\lambda) R_{N-1,h}(\lambda) 
\cdots  L_{1,h}(\lambda).
$$
The eq. (\ref{fyb}) guarantees the commutativity of
$I(\lambda)$ and the trace $T_a(\l)=\hbox {tr}_{\n} M_a (\l)$,
see the formula (\ref{tt}).

If we write the generators in terms of Weyl pair (\ref{wp})
$$
e=v u^{-1},
\hsp
k=u^2,
$$
$$
\tilde e=v u^{-3},
\hsp
\tilde k=u^2
$$
we get the expression for $R_f(\l)$
$$
R_f(\l)=\exp_{q^{-2}}
\left( (q-q^{-1})^{-1} v u^{-1}\otimes u^3 v^{-1}  \l
\right)\cdot
\exp_{q^{-2}}\left( (q-q^{-1})^{-1}
u v^{-1}\otimes v u^{-3}\l \right)\cdot
q^{{\frac{h_\a\otimes h_\a}{2}}}
$$
And after normalization one gets the formula for \fmt
\bn
r(\l,w)= R_f(1)^{-1} R_f(\l)=
\frac{\hbox{exp}_{q^{-2}} ((q-q^{-1})^{-1}w\l) 
\hbox{exp}_{q^{-2}}((q-q^{-1})^{-1}w^{-1}\l)}
{\hbox{exp}_{q^{-2}}((q-q^{-1})^{-1}w)
\hbox{exp}_{q^{-2}}((q-q^{-1})^{-1}w^{-1})}
\label{sovp}
\ed
where $w=vu\otimes uv^{-1}$.

In fact, the function $r(\l,w)$ satisfies the YB eq.
$$
r(\l,w_{1 2})r(\l\m,w_{2 3})r(\m,w_{1 2})=
r(\m,w_{2 3})r(\l\m,w_{1 2})r(\l,w_{2 3})
$$
with $w_{1 2}=vu\otimes uv^{-1}\otimes 1$ and
$w_{2 3}=1\otimes vu\otimes uv^{-1}$.

The following solution of this eq. has been found in \cite{FZ}
$$ 
r(\l,w) = 1 + \sum^{\infty}_{k=1}  
\frac{(1 - \l)(q -\l q^{-1})\ldots (q^{k-1} - \l q^{-k+1})}
{(q^{-1} -\l q)\ldots (q^{-k} - \l q^k)} w^k
$$

In \cite{V} it was noticed that r-matrix $r(\l,w)$
satisfies the functional eq.
$$
\frac{r(\l,wq)}{r(\l,wq^{-1})}=\frac{1+ \l w}{\l+w}
$$
To prove this fact it is sufficient to represent
q-exponent as an infinite product
$$
\hbox{exp}_{q^{-2}} ((q-q^{-1})^{-1}w)=\prod\limits_{k=0}^\infty (1+ q^{2k+1} w)
$$
Finally, multiplicative form (\ref{sovp}) was found in \cite{FV}.

One can write the standard form of YB eq. for \fmt
$P\cdot r(\l)$ (\ref{sovp}) 
and for Volterra L-operator (\ref{vlop})
\footnotemark
\footnotetext[3]
{P is the permutation operator: $P\cdot w_1\otimes w_2=  w_2\otimes w_1$}
$$
L_{n_2,a}(\l) L_{n_1,a}(\mu) P \cdot r(\mu/\l)=
P\cdot r(\mu/\l) L_{n_1,a}(\mu) L_{n_2,a}(\l)
$$

\section{Discussion and Concluding Remarks}
In this paper we explicitly demonstrated the
role of the \Rmt in the formalism of QISM 
on the example of Quantum Volterra model.

Starting from the \Rmt $\RMAT$ we found that the quantum \rep
for Volterra Model coincides with q-oscillator \rep. In this way
we derived Volterra L-operator and \fmt. The degenerated
q-oscillator \rep is the simplest for $\bp$ algebra.

It is interesting to generalize the one-to-one
correspondence between Integrable Model
and quantum \rep for \Rmt from q-oscillators to 
other \reps of $\bp$.

For example, let $V_\mu$  be Verma module for $\s$
with the highest weight $\mu \in  \Bbb {C}$. Then we denote
the evaluation \rep of the affine $\sl$ algebra as  $V_\mu(\l)$.  
Quantum L-operator derived with this \rep from \Rmt is
$$
L(\l)=
 \left( \begin{array}{cc}
k_\a^\half-\l^2 q^{-1}k_\a^{-\half} & -\l e_{-\a} k_\a^{-\half}(q-q^{-1})\\
\l e_\a k_\a^\half(q-q^{-1})  &  k_\a^{-\half}-\l^2 q^{-1}k_\a^{\half}
\end{array}
\right) 
$$
where 
$ e_{\pm \alpha} $ and $k_\a^{\pm 1}$ are  $\s$  generators. This L-operator
reminds one for the Lattice Sine-Gordon Model \cite{IK}.


\end{document}